\documentclass[twocolumn,A4]{article}
\usepackage[dvips]{graphics}
\usepackage{setspace}
\usepackage{color}
\definecolor{gold}{rgb}{0.85,0.66,0}
\definecolor{dblue}{rgb}{0,0,0.8}
\begin{document}
\onecolumn
\begin{center}
{\bf{\Large {\textcolor{gold}{Quantum transport in a mesoscopic ring: 
Evidence of an OR gate}}}}\\
~\\
{\textcolor{dblue}{Santanu K. Maiti}}$^{1,2,*}$ \\
~\\
{\em $^1$Theoretical Condensed Matter Physics Division,
Saha Institute of Nuclear Physics, \\
1/AF, Bidhannagar, Kolkata-700 064, India \\
$^2$Department of Physics, Narasinha Dutt College,
129, Belilious Road, Howrah-711 101, India} \\
~\\
{\bf Abstract}
\end{center}
We explore OR gate response in a mesoscopic ring threaded by a magnetic
flux $\phi$. The ring is symmetrically attached to two semi-infinite 
one-dimensional metallic electrodes and two gate voltages, viz, $V_a$ and 
$V_b$, are applied in one arm of the ring which are treated as the two 
inputs of the OR gate. All the calculations are based on the tight-binding 
model and the Green's function method, which numerically compute the 
conductance-energy and current-voltage characteristics as functions of 
the gate voltages, ring-to-electrodes coupling strengths and magnetic flux. 
Our theoretical study shows that, for $\phi=\phi_0/2$ ($\phi_0=ch/e$, 
the elementary flux-quantum) a high output current ($1$) (in the logical 
sense) appears if one or both the inputs to the gate are high ($1$), 
while if neither input is high ($1$), a low output current ($0$) appears. 
It clearly demonstrates the OR gate behavior and this aspect may be 
utilized in designing the electronic logic gate. 

\vskip 1cm
\begin{flushleft}
{\bf PACS No.}: 73.23.-b; 73.63.Rt. \\
~\\
{\bf Keywords}: A. Mesoscopic ring; D. Conductance; D. $I$-$V$ 
characteristic; D. OR gate.
\end{flushleft}
\vskip 4.2in
\noindent
{\bf ~$^*$Corresponding Author}: Santanu K. Maiti

Electronic mail: santanu.maiti@saha.ac.in

\newpage
\twocolumn

\section{Introduction}

The study of electron transport in quantum confined model system is a
challenging field in the modern age of nanoscience and technology, 
since these simple looking systems can be used to design nanodevices
especially in electronic as well as spintronic engineering. A mesoscopic
metallic ring is a nice example of quantum confined systems, and, with 
the help of the ring we can make a device that can act as a logic 
gate, which may be used in nanoelectronic circuits. To reveal this
phenomenon we design a bridge system where the ring is attached to
two external electrodes, the so-called electrode-ring-electrode bridge.
Based on the pioneering work of Aviram and Ratner~\cite{aviram}, the
theoretical description of electron transport in a bridge system has got 
much progress. Later, many excellent experiments~\cite{reed1,reed2,tali} 
have been done in several bridge systems to understand the basic 
mechanisms underlying the electron transport. Though in literature both 
theoretical~\cite{nitzan1,nitzan2,peeters1,peeters2,orella1,orella2,new,
muj1,muj2,walc2,walc3,cui} as well as experimental~\cite{reed1,reed2,tali} 
works on electron transport are available, yet lot of controversies are 
still present between the theory and experiment, and the complete knowledge 
of the conduction mechanism in this scale is not very well established 
even today. The ring-electrodes interface structure significantly 
controls the electronic transport in the ring. By changing the 
geometry, one can tune the transmission probability of an electron 
across the ring which is solely due to the effect of quantum interference 
among the electronic waves passing through different arms of the ring. 
Furthermore, the electron transport in the ring can be modulated in 
other way by tuning the magnetic flux, the so-called Aharonov-Bohm (AB) 
flux, that threads the ring. The AB flux threading the ring may change 
the phases of the wave functions propagating along the different arms 
of the ring leading to constructive or destructive interferences, and 
therefore, the transmission amplitude changes~\cite{baer2,baer3,tagami,
walc1,baer1}. Beside these factors, ring-to-electrodes coupling is 
another important issue that controls the electron transport in a 
meaningful way~\cite{baer1}. All these are the key factors which 
regulate the electron transmission in the electrode-ring-electrode 
bridge system and these effects have to be taken into account properly 
to reveal the transport mechanisms. 

Our main focus of the present paper is to describe the OR gate response in a 
mesoscopic ring threaded by a magnetic flux $\phi$. The ring is contacted 
symmetrically to the electrodes, and the two gate voltages $V_a$ and $V_b$
are applied in one arm of the ring (see Fig.~\ref{or}) those are treated 
as the two inputs of the OR gate. Here we adopt a simple tight-binding 
model to describe the system and all the calculations are performed 
numerically. We address the OR gate behavior by studying the 
conductance-energy 
and current-voltage characteristics as functions of the ring-electrodes 
coupling strengths, magnetic flux and gate voltages. Our study reveals 
that for a particular value of the magnetic flux, $\phi=\phi_0/2$, a high 
output current ($1$) (in the logical sense) is available if one or both
the inputs to the gate are high ($1$), while if neither input is high ($1$),
a low output current ($0$) appears. This phenomenon clearly demonstrates 
the OR behavior. To the best of our knowledge the OR gate response 
in such a simple system has not been addressed earlier in the literature.

We organize the paper as follow. Following the introduction 
(Section $1$), in Section $2$, we present the model and the theoretical 
formulations for our calculations. Section $3$ discusses the results, 
and finally, we summarize our results in Section $4$.

\section{Model and the theoretical background}

\begin{figure}[ht]
{\centering \resizebox*{7cm}{3.8cm}{\includegraphics{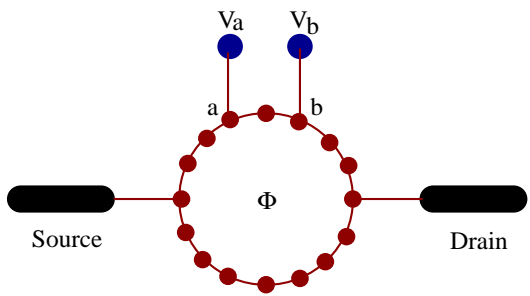}}\par}
\caption{(Color online). A mesoscopic ring with total number of atomic
sites $N=16$ (filled red circles), threaded by a magnetic flux 
$\phi$, is attached to $1$D metallic electrodes, viz, source and drain. 
The atoms $a$ and $b$ are subjected to the gate voltages $V_a$ and $V_b$ 
respectively, those are variable.}
\label{or}
\end{figure}

We start by referring to Fig.~\ref{or}. A mesoscopic ring, threaded by a
magnetic flux $\phi$, is attached symmetrically (upper and lower arms have
equal number of lattice points) to two semi-infinite one-dimensional 
($1$D) metallic electrodes. The atoms $a$ and $b$ in the upper arm of the 
ring are subjected to the gate voltages $V_a$ and $V_b$ respectively, and 
these are treated as the two inputs of the OR gate.

Using the Landauer conductance formula~\cite{datta,marc} we calculate the
conductance ($g$) of the ring. At very low temperature and bias voltage it 
can be expressed in the form,
\begin{equation}
g=\frac{2e^2}{h} T
\label{equ1}
\end{equation}
where $T$ gives the transmission probability of an electron through the ring. 
This $(T)$ can be represented in terms of the Green's function of the 
ring and its coupling to the two electrodes by the 
relation~\cite{datta,marc},
\begin{equation}
T={\mbox{Tr}} \left[\Gamma_S G_{R}^r \Gamma_D G_{R}^a\right]
\label{equ2}
\end{equation}
where $G_{R}^r$ and $G_{R}^a$ are respectively the retarded and advanced
Green's functions of the ring including the effects of the electrodes.
The parameters $\Gamma_S$ and $\Gamma_D$ describe the coupling of the
ring to the source and drain respectively. For the full system i.e., 
the ring, source and drain, the Green's function is defined as,
\begin{equation}
G=\left(E-H\right)^{-1}
\label{equ3}
\end{equation}
where $E$ is the injecting energy of the source electron. To Evaluate
this Green's function, the inversion of an infinite matrix is needed since
the full system consists of the finite ring and the two semi-infinite 
electrodes. However, the entire system can be partitioned into sub-matrices 
corresponding to the individual sub-systems and the Green's function for 
the ring can be effectively written as,
\begin{equation}
G_{R}=\left(E-H_{R}-\Sigma_S-\Sigma_D\right)^{-1}
\label{equ4}
\end{equation}
where $H_{R}$ is the Hamiltonian of the ring that can be expressed within 
the non-interacting picture like,
\begin{eqnarray}
H_{R} & = & \sum_i \left(\epsilon_{i0} + V_a \delta_{ia} + V_b \delta_{ib} 
\right) c_i^{\dagger} c_i \nonumber \\
 & + & \sum_{<ij>} t \left(c_i^{\dagger} c_j e^{i\theta}+ c_j^{\dagger} 
c_i e^{-i\theta}\right)
\label{equ5}
\end{eqnarray}
In this Hamiltonian $\epsilon_{i0}$'s are the site energies for all the 
sites $i$ except the sites $i=a$ and $b$ where the gate voltages $V_a$ 
and $V_b$ are applied, those are variable. These gate voltages can be 
incorporated through the site energies as expressed in the above 
Hamiltonian. $c_i^{\dagger}$ ($c_i$) is the creation (annihilation) 
operator of an electron at the site $i$ and $t$ is the nearest-neighbor 
hopping integral. The phase factor $\theta=2 \pi \phi/N \phi_0$ comes 
due to the flux $\phi$ threaded by the ring, where $N$ corresponds to 
the total number of atomic sites in the ring. Similar kind of 
tight-binding Hamiltonian is also used, except the phase factor 
$\theta$, to describe the $1$D perfect electrodes where the 
Hamiltonian is parametrized by constant on-site potential $\epsilon_0$ 
and nearest-neighbor hopping integral $t_0$. The hopping integral between
the source and the ring is $\tau_S$, while it is $\tau_D$ between the
ring and the drain. The parameters $\Sigma_S$ and $\Sigma_D$ 
in Eq.~(\ref{equ4}) represent the self-energies due to the coupling of 
the ring to the source and drain respectively, where all the informations 
of this coupling are included into these self-energies.

To evaluate the current ($I$), passing through the ring, as a function 
of the applied bias voltage ($V$) we use the relation~\cite{datta},
\begin{equation}
I(V)=\frac{e}{\pi \hbar}\int \limits_{E_F-eV/2}^{E_F+eV/2} T(E)~ dE
\label{equ8}
\end{equation}
where $E_F$ is the equilibrium Fermi energy. Here we assume that the 
entire voltage is dropped across the ring-electrode interfaces, and it 
is examined that under such an assumption the $I$-$V$ characteristics 
do not change their qualitative features. 

All the results in this presentation are computed at absolute
zero temperature, but they should valid even for finite temperature
($\sim 300$ K) as the broadening of the energy levels of the ring due
to its coupling with the electrodes becomes much larger than that of
the thermal broadening~\cite{datta}. For simplicity, we take the unit 
$c=e=h=1$ in our present calculations. 

\section{Results and discussion}

To illustrate the results, let us first mention the values of the different
parameters used for the numerical calculations. In the ring, the on-site
energy $\epsilon_{i0}$ is taken as $0$ for all the sites $i$, except the
sites $i=a$ and $b$ where the site energies are taken as $V_a$ and $V_b$ 
respectively, and the nearest-neighbor hopping strength $t$ is set
to $3$. On the other hand, for the side attached electrodes the 
on-site energy ($\epsilon_0$) and the nearest-neighbor hopping strength 
($t_0$) are fixed to $0$ and $4$ respectively. Throughout the study, we 
focus our results for the two limiting cases depending on the strength 
of the coupling of the ring to the source and drain. In one case we use 
the condition $\tau_{S(D)} << t$, which is the so-called weak-coupling 
limit. For this regime we choose $\tau_S=\tau_D=0.5$. In the other case
the condition $\tau_{S(D)} \sim t$ is used, which is named as the 
strong-coupling limit. In this particular regime, the values of the 
parameters are set as $\tau_S=\tau_D=2.5$. The key controlling 
parameter for all these calculations is the magnetic flux $\phi$ 
which is set to $\phi_0/2$ i.e., $0.5$ in our chosen unit.

\begin{figure}[ht]
{\centering \resizebox*{8cm}{7cm}{\includegraphics{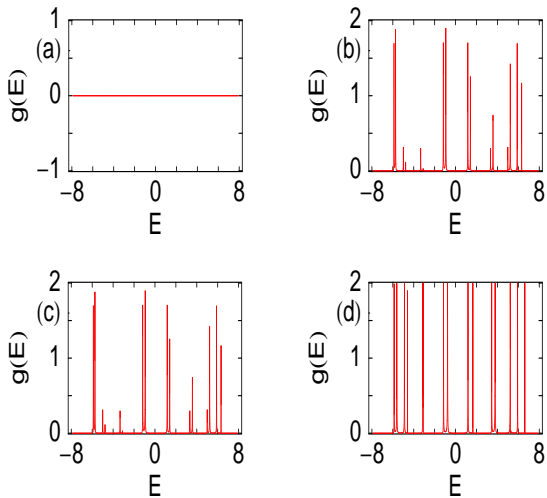}}\par}
\caption{(Color online). $g$-$E$ characteristics for a mesoscopic ring 
with $N=16$ and $\phi=0.5$ in the weak-coupling limit.
(a) $V_a=V_b=0$, (b) $V_a=2$ and $V_b=0$, (c) $V_a=0$ and $V_b=2$ and
(d) $V_a=V_b=2$.}
\label{condlow}
\end{figure}

As illustrative examples, in Fig.~\ref{condlow} we display the 
conductance-energy ($g$-$E$) characteristics for a mesoscopic ring 
considering $N=16$ in the limit of weak ring-to-electrodes coupling, 
where (a), (b), (c) and (d) correspond to the results for the four 
different cases of the gate voltages $V_a$ and $V_b$. In the particular
case when $V_a=V_b=0$ i.e., both the inputs are low ($0$), the conductance
shows the value $0$ in the entire energy range (Fig.~\ref{condlow}(a)). 
This clearly indicates that the electron cannot conduct from the source
to the drain across the ring. While, for the other three cases
i.e., $V_a=2$ and $V_b=0$ (Fig.~\ref{condlow}(b)), $V_a=0$ and $V_b=2$
(Fig.~\ref{condlow}(c)) and $V_a=V_b=2$ (Fig.~\ref{condlow}(d)), the
conductance shows fine resonance peaks for some particular energies.  
Thus, in all these three cases, the electron can conduct through the
ring. From Fig.~\ref{condlow}(d) it is observed that at the resonances
the conductance $g$ approaches the value $2$, and accordingly, the 
transmission probability $T$ goes to unity, since the relation $g=2T$
is satisfied from the Landauer conductance formula (see Eq.~\ref{equ1}
with $e=h=1$). On the other hand, the transmission probability $T$
decays from $1$ for the cases where any one of the two gate voltages 
is high and other is low (Figs.~\ref{condlow}(b) and (c)). All these
resonance peaks are associated with the energy eigenvalues of the 
ring, and therefore, we can say that the conductance spectrum reveals
itself the electronic structure of the ring. Now we interpret the
dependences of the two gate voltages in these four different cases.
The probability amplitude of getting an electron across the ring depends 
on the quantum interference of the electronic waves passing through the 
two arms of the ring. For the ring attached symmetrically to the electrodes
i.e., when the two arms of the ring are identical with each other, the 
probability amplitude is exactly zero ($T=0$) for the flux $\phi=\phi_0/2$. 
This is due to the result of the quantum interference among the two waves 
in the two arms of the ring, which can be obtained in a very simple 
mathematical calculation. Thus for the cases when both the two inputs
($V_a$ and $V_b$) are zero (low), the two arms of the ring become identical,
and therefore, the transmission probability drops to zero. On the other 
hand, for the other three cases the symmetry of the two arms of the ring 
\begin{figure}[ht]
{\centering \resizebox*{8cm}{7cm}{\includegraphics{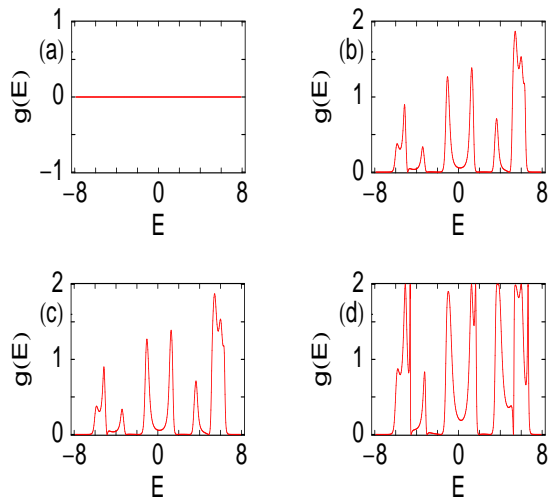}}\par}
\caption{(Color online). $g$-$E$ characteristics for a mesoscopic ring 
with $N=16$ and $\phi=0.5$ in the strong-coupling limit.
(a) $V_a=V_b=0$, (b) $V_a=2$ and $V_b=0$, (c) $V_a=0$ and $V_b=2$ and
(d) $V_a=V_b=2$.}
\label{condhigh}
\end{figure}
is broken when either the atom $a$ or $b$ or the both are subjected to
the gate voltages $V_a$ and $V_b$, and therefore, the non-zero value 
of the transmission probability is achieved which reveals the electron 
conduction across the ring. The reduction of the transmission probability
for the cases when any one of the two gates is high and other is low
compared to the case where both the gates are high is also due to the 
quantum interference effect. Thus we can predict that the electron 
conduction takes place across the ring if any one or both the inputs 
to the gate are high, while if both the inputs are low, the conduction 
is no longer possible. This feature clearly demonstrates the OR gate 
behavior. In the limit of strong-coupling, we get additional one feature
compared to those as presented in the limit of weak-coupling. The results
are shown in Fig.~\ref{condhigh}, where the results in (a), (b), (c) and
(d) are drawn respectively for the same gate voltages as in 
Fig.~\ref{condlow}. For the strong-coupling limit, all the resonances
get substantial widths compared to the weak-coupling limit. The 
contribution for the broadening of the resonance peaks in this 
strong-coupling limit appears from the imaginary parts of the 
self-energies $\Sigma_S$ and $\Sigma_D$ respectively~\cite{datta}. 
Hence by tuning the coupling strength, we can get the electron 
transmission across the ring for the wider range of energies and 
it provides an important signature in the study of current-voltage 
($I$-$V$) characteristics.

All these features of electron transfer become much more clearly visible
by studying the $I$-$V$ characteristics. The current passing through the
\begin{figure}[ht]
{\centering \resizebox*{8cm}{7cm}{\includegraphics{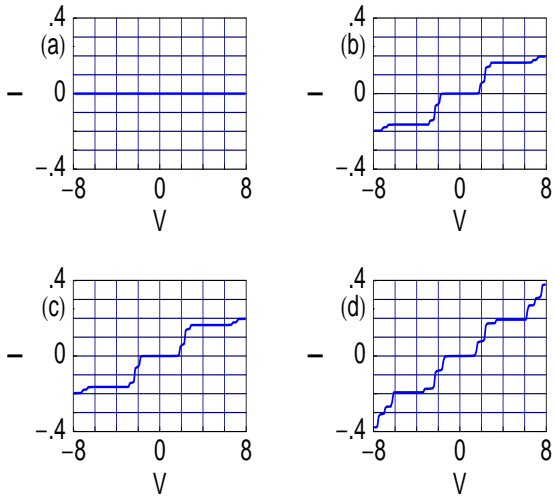}}\par}
\caption{(Color online). $I$-$V$ characteristics for a mesoscopic ring 
with $N=16$ and $\phi=0.5$ in the weak-coupling limit. (a) $V_a=V_b=0$, 
(b) $V_a=2$ and $V_b=0$, (c) $V_a=0$ and $V_b=2$ and (d) $V_a=V_b=2$.}
\label{currlow}
\end{figure}
ring is computed from the integration procedure of the transmission function
$T$ as prescribed in Eq.~\ref{equ8}. The transmission function varies exactly
similar to that of the conductance spectrum, differ only in magnitude by the
factor $2$ since the relation $g=2T$ holds from the Landauer conductance
formula Eq.~\ref{equ1}. As representative examples, in Fig.~\ref{currlow}
we display the current-voltage characteristics for a mesoscopic ring with
$N=16$ in the limit of weak-coupling. For the case when both the two inputs
are zero, the current $I$ is zero (see Fig.~\ref{currlow}(a)) for the 
\begin{table}[ht]
\begin{center}
\caption{OR gate behavior in the limit of weak-coupling. The current 
$I$ is computed at the bias voltage $6.02$.}
\label{table1}
~\\
\begin{tabular}{|c|c|c|}
\hline \hline
Input-I ($V_a$) & Input-II ($V_b$) & Current ($I$) \\ \hline 
$0$ & $0$ & $0$ \\ \hline
$2$ & $0$ & $0.164$ \\ \hline
$0$ & $2$ & $0.164$ \\ \hline
$2$ & $2$ & $0.193$ \\ \hline \hline
\end{tabular}
\end{center}
\end{table}
entire bias voltage $V$. This feature is clearly visible from the 
conductance spectrum, Fig.~\ref{condlow}(a), since the current is 
computed from the integration procedure of the transmission function 
$T$. In the rest three cases, a high output current is obtained those
are clearly presented in Figs.~\ref{currlow}(b), (c) and (d). From
these figures it is observed that
\begin{figure}[ht]
{\centering \resizebox*{8cm}{7cm}{\includegraphics{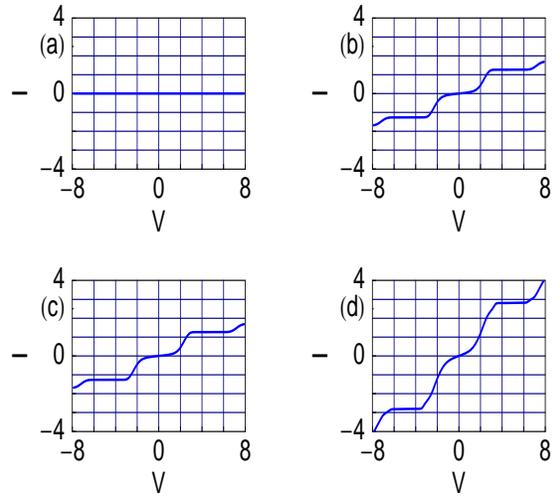}}\par}
\caption{(Color online). $I$-$V$ characteristics for a mesoscopic ring 
with $N=16$ and $\phi=0.5$ in the strong-coupling limit. (a) $V_a=V_b=0$, 
(b) $V_a=2$ and $V_b=0$, (c) $V_a=0$ and $V_b=2$ and (d) $V_a=V_b=2$.}
\label{currhigh}
\end{figure}
the current exhibits staircase-like structure with fine steps as a
function of the applied bias voltage. This is due to the existence of
the sharp resonance peaks in the conductance spectrum in the weak-coupling
limit, since the current is computed by the integration method of the
transmission function $T$. With the increase of the bias voltage $V$,
the electrochemical potentials on the electrodes are shifted gradually,
and finally cross one of the quantized energy levels of the ring.
Accordingly, a current channel is opened up which provides a jump
in the $I$-$V$ characteristic curve. Addition to these behaviors, it is
also important to note that the non-zero value of the current appears
beyond a finite value of $V$, the so-called threshold voltage ($V_{th}$).
This $V_{th}$ can be controlled by tuning the size ($N$) of the ring.
From these $I$-$V$ characteristics, the OR gate response is understood
very easily. To make it much clear, in Table~\ref{table1}, we show a 
quantitative estimate of the typical current amplitude determined at the
bias voltage $V=6.02$, in this limit of weak ring-to-electrodes coupling.
\begin{table}[ht]
\begin{center}
\caption{OR gate behavior in the limit of strong-coupling. The current 
$I$ is computed at the bias voltage $6.02$.}
\label{table2}
~\\
\begin{tabular}{|c|c|c|}
\hline \hline
Input-I ($V_a$) & Input-II ($V_b$) & Current ($I$) \\ \hline 
$0$ & $0$ & $0$ \\ \hline
$2$ & $0$ & $1.268$ \\ \hline
$0$ & $2$ & $1.268$ \\ \hline
$2$ & $2$ & $2.824$ \\ \hline \hline
\end{tabular}
\end{center}
\end{table}
It is observed that when any one of the two gates is high and other is low,
the current gets the value $0.164$, and for the case when both the two inputs
are high, it ($I$) achieves the value $0.193$. While, for the case when
both the two inputs are low ($V_a=V_b=0$), the current becomes exactly 
zero. In the same analogy, as above, here we also discuss the $I$-$V$
characteristics for the strong-coupling limit. In this limit, the
current varies almost continuously with the applied bias voltage and
achieves much larger amplitude than the weak-coupling case as presented
in Fig.~\ref{currhigh}. The reason is that, in the limit of strong-coupling
all the energy levels get broadened which provide larger current in the
integration procedure of the transmission function $T$. Thus by tuning the
strength of the ring-to-electrodes coupling, we can achieve very large
current amplitude from the very low one for the same bias voltage $V$.
All the other properties i.e., the dependences of the gate voltages on the
$I$-$V$ characteristics are exactly similar to those as given in
Fig.~\ref{currlow}. In this strong-coupling limit we also make a quantitative
study for the typical current amplitude, given in Table~\ref{table2}, where
the current amplitude is determined at the same bias voltage ($V=6.02$) as
earlier. The response of the output current is exactly similar to that as
given in Table~\ref{table1}. Here the current achieves the value $1.268$ in
the cases where any one of the two gates is high and other is low, and it
($I$) becomes $2.824$ for the case where both the two inputs are high. On
the other hand, the current becomes exactly zero for the case
where $V_a=V_b=0$. The non-zero values of the current in this 
strong-coupling limit are much larger than the weak-coupling case, as 
expected. From these results we can clearly manifest that a mesoscopic 
ring exhibits the OR gate response.

\section{Concluding remarks}

To summarize, we have addressed the OR gate response in a mesoscopic 
metallic ring threaded by a magnetic flux $\phi$ and attached 
symmetrically to the electrodes. Two atoms in the upper arm of the 
ring are subjected to the gate voltages $V_a$ and $V_b$ respectively 
those are taken as the two inputs of the OR gate. The system is described 
by the tight-binding Hamiltonian and all the calculations are done in 
the Green's function formalism. We have numerically computed the 
conductance-energy and current-voltage characteristics as functions 
of the gate voltages, ring-to-electrodes coupling strengths and 
magnetic flux. Very interestingly we have noticed that, for the half 
flux-quantum value of $\phi$ ($\phi=\phi_0/2$), a high output current 
($1$) (in the logical sense) appears if one or both the inputs to 
the gate are high ($1$). On the other hand, if neither input is high 
($1$), a low output current ($0$) appears. It clearly manifests the 
OR behavior and this aspect may be utilized in designing a tailor 
made electronic logic gate. 

In this presentation, we have explored the conductance-energy and 
current-voltage characteristics for some fixed parameter values 
considering a ring with total number of atomic sites $N=16$. Though the 
results presented here change numerically for the other parameter 
values and ring size ($N$), but all the basic features remain 
exactly invariant. 

The importance of this article is concerned with (i) the simplicity of the
geometry and (ii) the smallness of the size. To the best of our knowledge 
the OR gate response in such a simple low-dimensional system that can be 
operated even at finite temperature ($\sim 300$ K) has not been addressed 
in the literature.

\end{document}